\documentclass[12pt,a4paper,dvips]{article}
\usepackage{a4p}
\usepackage{cite,mcite}
\usepackage{graphicx}
\usepackage{physics }
\usepackage{l3_title,ifthen}
\usepackage{rotating}
\journalname{Phys. Lett. B}

\usepackage{Lep}
\Lep{2}

\date{December 17, 1999}
\preprint{99-180}

\newlength{\capindent}
\newlength{\capwidth}
\newlength{\figwidth}
\newcommand{\icaption}[2][!*!,!]{\hspace*{\capindent}%
  \begin{minipage}{\capwidth}
    \ifthenelse{\equal{#1}{!*!,!}}%
      {\caption{#2}}%
      {\caption[#1]{#2}}
  \end{minipage}}

\begin{document}
\begin{titlepage}
\title{Hard-Photon Production and Tests of QED at LEP} 
\author{The L3 Collaboration}
%
%
\begin{abstract}
\indent
The total and differential cross sections of the 
process $\epem \ra n \gamma$ with $n \geq 2$ are measured
using data collected by the L3 experiment 
at centre--of--mass energies of $\sqrt{s}=183$ and $189\GeV$. The 
results are in agreement with the 
Standard Model expectations. Limits are set on deviations
from QED, contact interaction cut-off parameters and masses of excited 
electrons.
\end{abstract}
\submitted

\end{titlepage}
%
%
\section{Introduction}                                         
\indent

The process $\epem \ra \gamma\gamma (\gamma)$, where $(\gamma)$
denotes possible additional photons, is described very
accurately by QED. The experimental signature of these events is clean, and 
they can be selected with negligible background. Therefore, this process is well 
suited to test QED and to look for new physics phenomena, whose expected 
contributions grow with the increase of the centre--of--mass energy, $\sqrt{s}$.

In this paper, the results on the study of the process 
$\epem \ra n \gamma$ ($n \geq 2$) are presented. The analysis is
performed on the data sample collected by the L3 detector~\cite{l3rf} during
1997 and 1998, at $\sqrt{s} =182.7\GeV$ ($183\GeV$ hereafter) and
$\sqrt{s} =188.7\GeV$ ($189\GeV$ hereafter) respectively. The
integrated luminosities for each sample are 54.8 pb$^{-1}$ and 175.3 pb$^{-1}$, 
respectively. Previous results have been published 
by L3 at lower centre-of-mass energies~\cite{gg91,gg130,gg172} and by
other experiments~\cite{ggother}. 

%
%
\section{Event selection}
\indent

The analysis performed on these data is similar to that reported in previous
papers~\cite{gg172}. A photon candidate is defined as:

\begin{itemize}
\item A shower in the electromagnetic barrel or end-cap calorimeters with energy 
      larger than 1 \GeV{}. The profile of the shower must be consistent with 
      that of an electromagnetic particle.

\item The number of hits in the vertex chamber within an azimuthal angle of
      $\pm 8^{\circ}$ around the direction of the photon candidate must be less than
      40\% of the expected number of hits for a charged particle. 
\end{itemize}

To select an event there must be at least two photon candidates with 
polar angles $\theta_{\gamma}$ between $16^{\circ}$ and $164^{\circ}$ 
with an angular separation of more than $15^{\circ}$ and no other activity
in the detector. In addition, to reject 
$\epem \ra \nu \bar{\nu} \gamma \gamma$ and cosmic rays, we require that the sum of the 
energies of the photon candidates be larger than $\sqrt{s}/2$.

The only expected backgrounds  are $\epem \ra \epem (\gamma)$ and
$\epem \ra \tau^{+} \tau^{-} (\gamma)$. These 
contributions are estimated from Monte Carlo simulations using 
BHWIDE~\cite{bhwide} for Bhabha events and KORALZ~\cite{koralz} for 
$\tau$ events, and are found to be negligible. The acceptance is computed 
applying the same analysis to a 
sample of $\epem \ra \gamma \gamma (\gamma)$ events generated using an 
order $\alpha^{3}$ Monte Carlo generator~\cite{ggg} 
passed through the L3 simulation~\cite{geant} and reconstruction 
programs. The selection efficiencies to detect at least two photons
inside the fiducial volume are found to 
be  $68.8 \pm 0.2 \%$ at $\sqrt{s}=183$ \GeV{} and
$68.0 \pm 0.2 \%$ at $\sqrt{s}=189$ \GeV{} for 
$16^{\circ} < \theta_{\gamma} < 164^{\circ}$, where the errors quoted 
are the statistical errors of the Monte Carlo sample. The
efficiency of the calorimetric energy trigger
is estimated to be above 99.7 \% for both samples. It is
estimated by using a sample of Bhabha events, which has an independent trigger
for charged particles.

%
%

\section{Analysis}
\indent

\begin{figure}
\begin{center}
\includegraphics[width=15.0truecm]{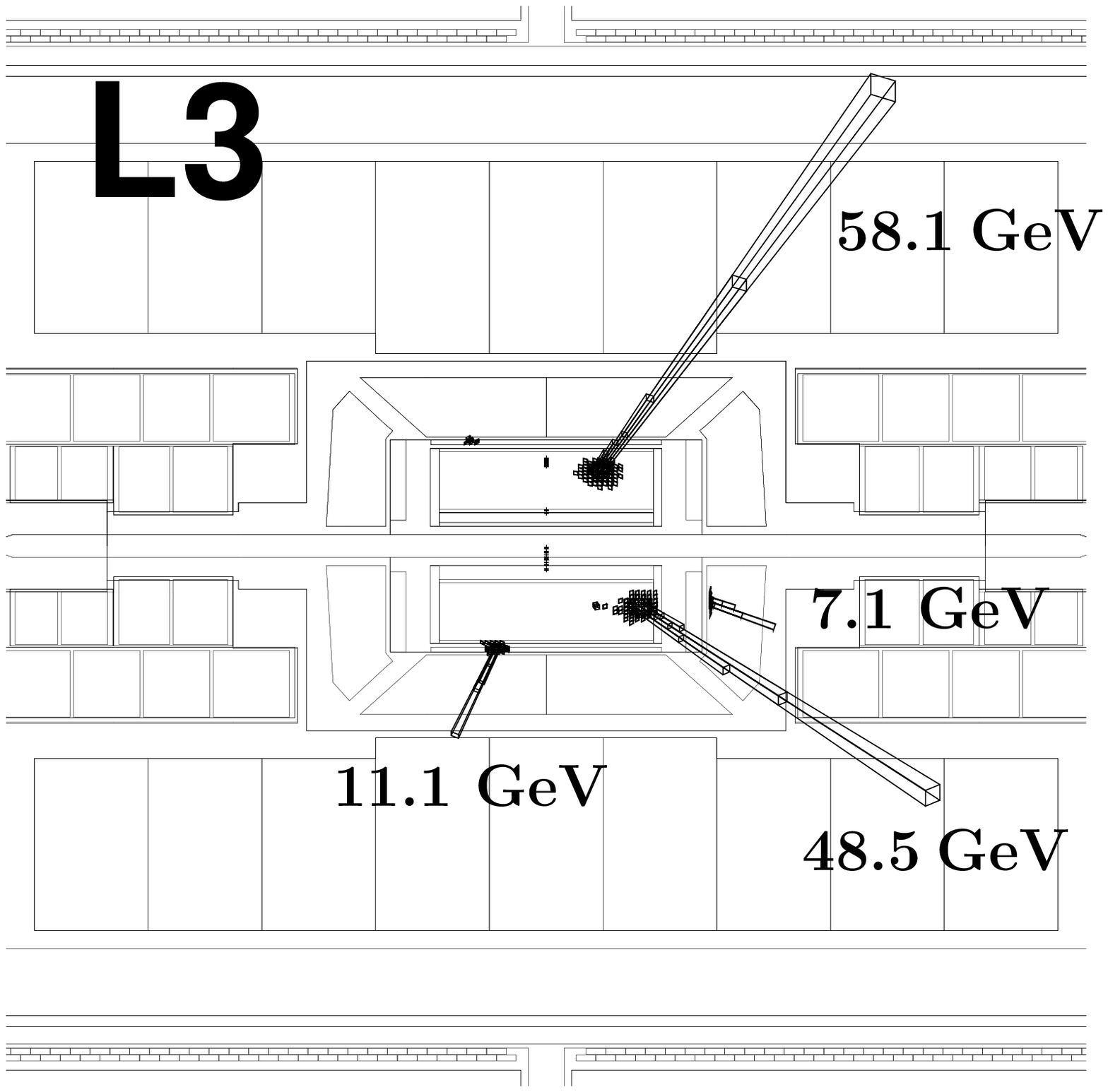}
\end{center}
\icaption{\label{fig:scan}
    Display of an event with four detected photons at $\sqrt{s}=183\GeV$.}
\end{figure}

\begin{figure}
\begin{center}
\includegraphics[width=10.0truecm]{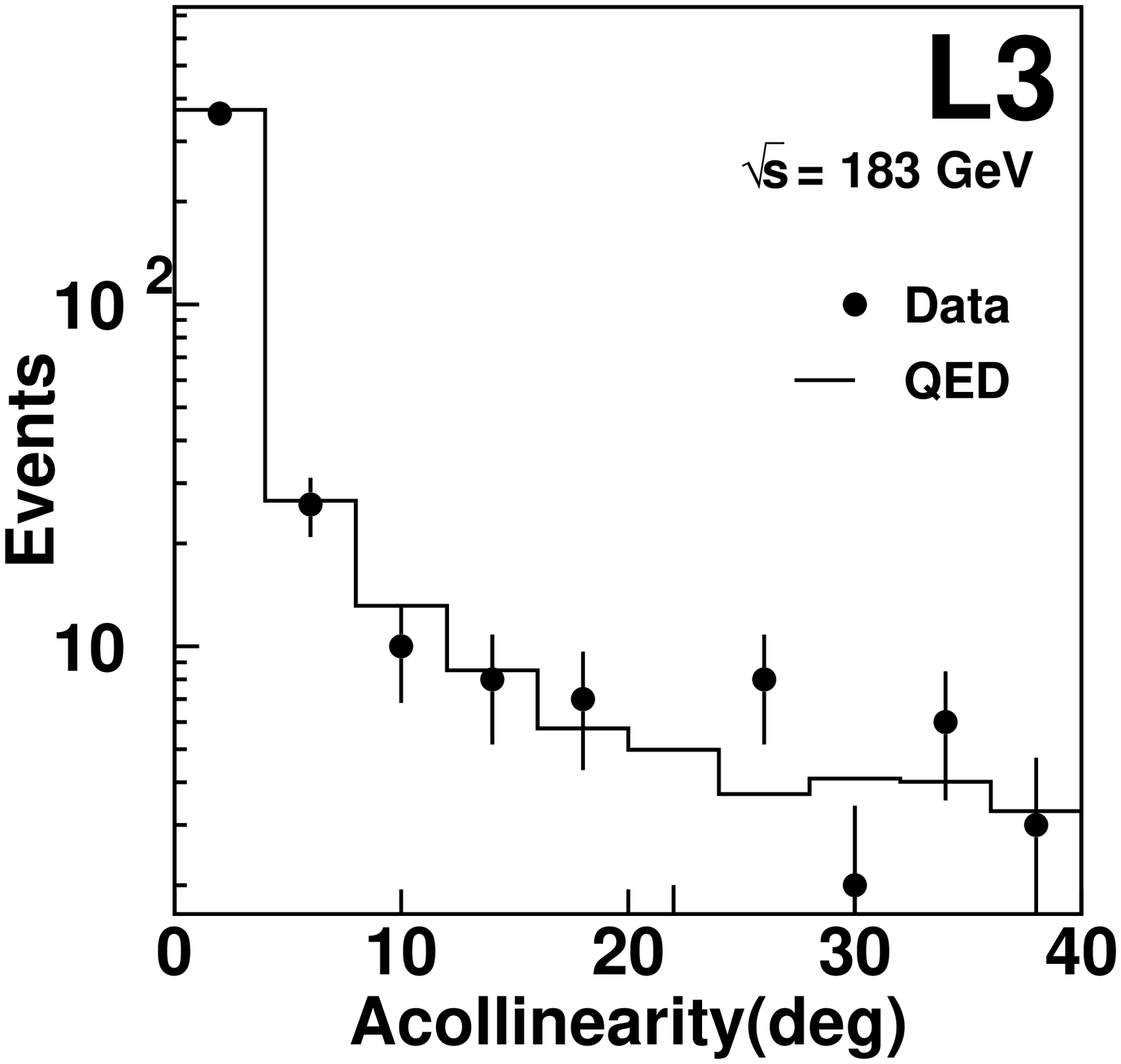}
\end{center}
\begin{center}
\includegraphics[width=10.0truecm]{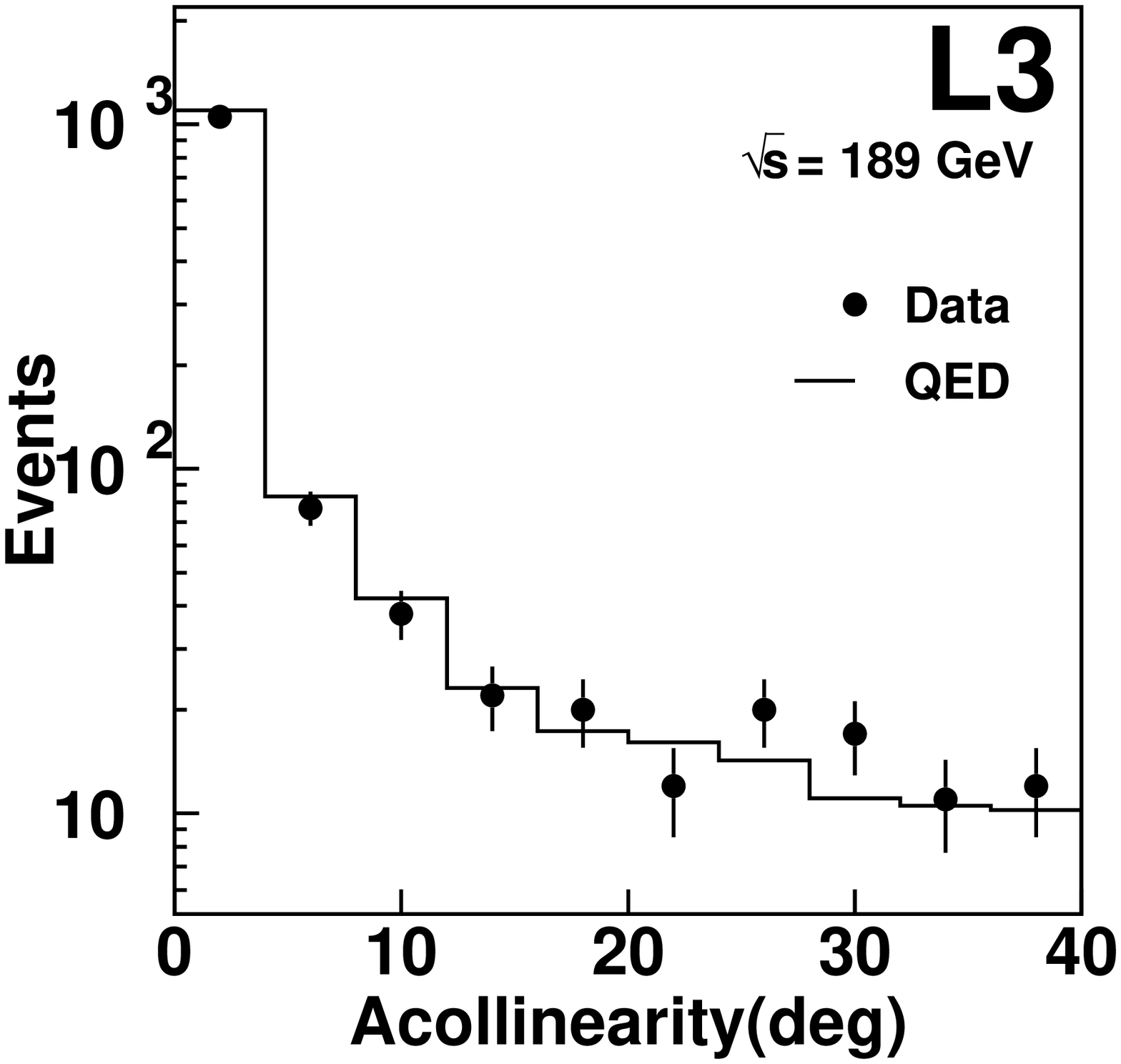}
\end{center}
\icaption{\label{fig:acol}
    Distribution of the acollinearity angle between the two most energetic 
    photons in the $\epem \ra \gamma\gamma(\gamma)$ process at 
    $\sqrt{s}=183\GeV$ (top), and $\sqrt{s}=189\GeV$ (bottom). }
\end{figure}

\begin{figure}
\begin{center}
\includegraphics[width=10.0truecm]{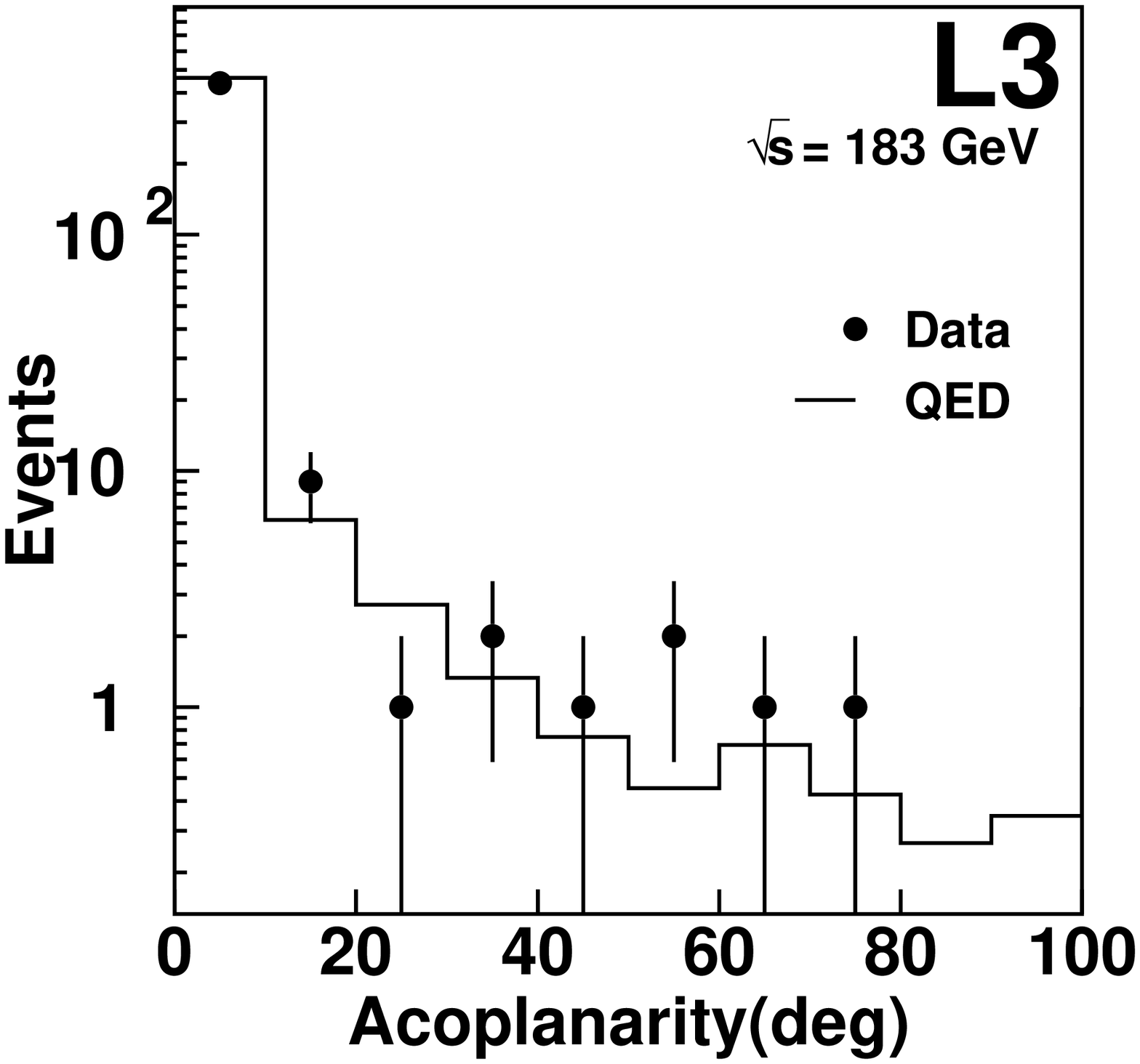}
\end{center}
\begin{center}
\includegraphics[width=10.0truecm]{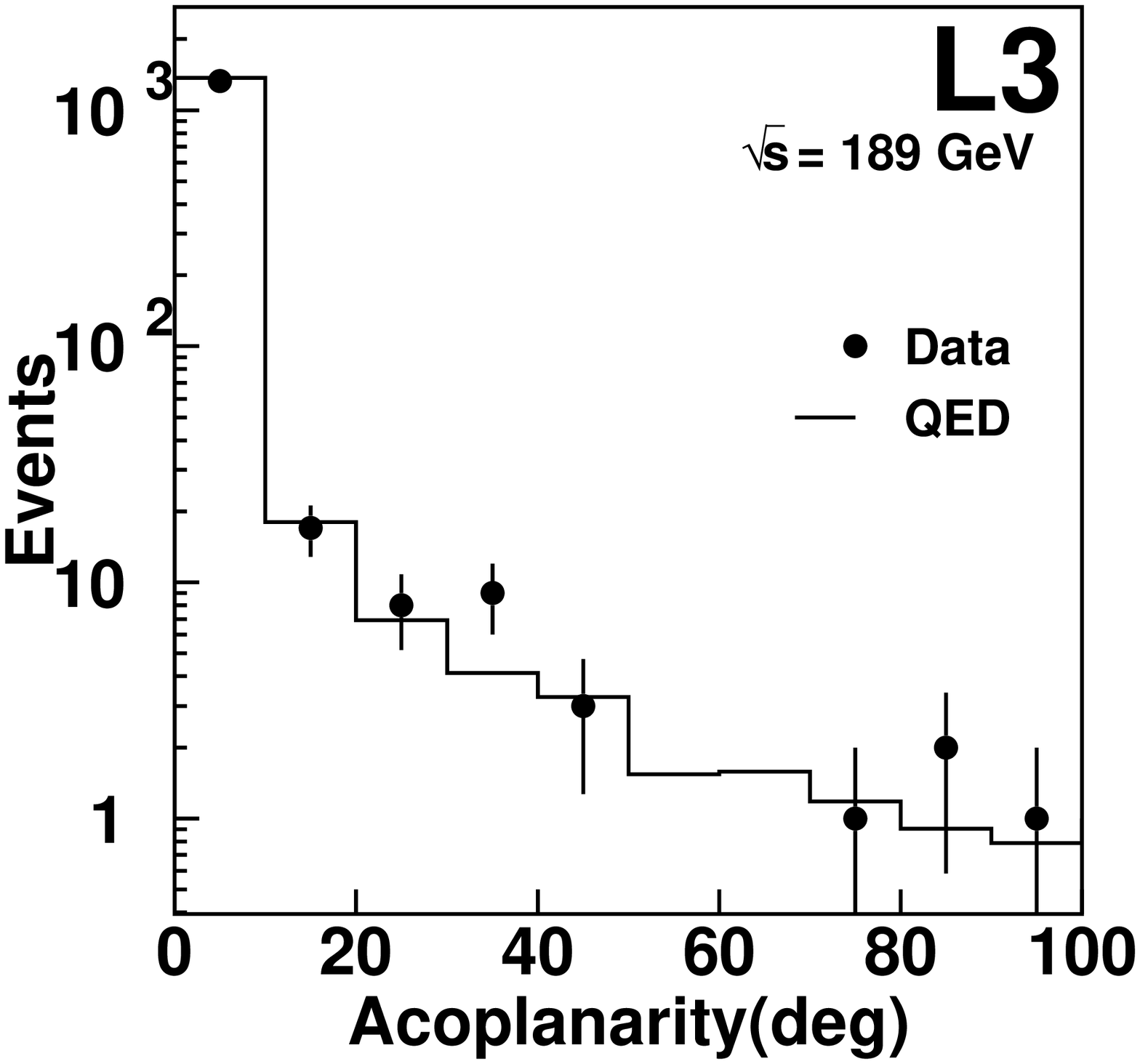}
\end{center}
\icaption{\label{fig:acopl}
    Distribution of the acoplanarity angle between the two most energetic 
    photons in the $\epem \ra \gamma\gamma(\gamma)$ process at 
    $\sqrt{s}=183\GeV$ (top), and $\sqrt{s}=189\GeV$ (bottom). } 
\end{figure}

A total of 460 events at $\sqrt{s}=183\GeV$ and 1374 events at $\sqrt{s}=189\GeV$ 
are selected. They are classified according to the number of isolated 
photons in $16^{\circ} < \theta_{\gamma} < 164^{\circ}$, as presented 
in Table~\ref{tab:number_of_g}, together with the number of
expected events. Figure~\ref{fig:scan} shows one event with 4 detected photons at $\sqrt{s}=183\GeV$. No events
with 5 or more photons in this angular range have been observed. For 
the two most energetic photons, the acollinearity angle distribution 
is shown in Figure~\ref{fig:acol}, and the acoplanarity angle distribution 
in Figure~\ref{fig:acopl}.

\begin{table}[h]
\begin{center}
\begin{tabular}{|l||c|c||c|c|}        
\cline{2-5} 
\multicolumn{1}{c|}{~} & \multicolumn{2}{c||}{$\sqrt{s}=183\GeV$ } &  \multicolumn{2}{c|}{$\sqrt{s}=189\GeV$ } \\ \cline{2-5}
\multicolumn{1}{c|}{~} & $N_{obs}$ &  $N_{exp}$       &  $N_{obs}$         &  $N_{exp}$     \\ \hline 
 $ 2{\gamma}$ & $   436    $       & $      453  $    & $  1302    $       & $     1346  $  \\
 $ 3{\gamma}$ & $    23    $       & $       24  $    & $    72    $       & $       69  $  \\
 $ 4{\gamma}$ & $     1    $       & $      0.04 $    & $     0    $       & $      0.1  $  \\ \hline
\end{tabular}
\end{center}

\icaption{Number of observed, $N_{obs}$, and expected, $N_{exp}$, events with 2, 3 and 4 photons.
          \label{tab:number_of_g}}
\end{table}                                                    

\begin{figure}
\begin{center}
\includegraphics[width=10.0truecm]{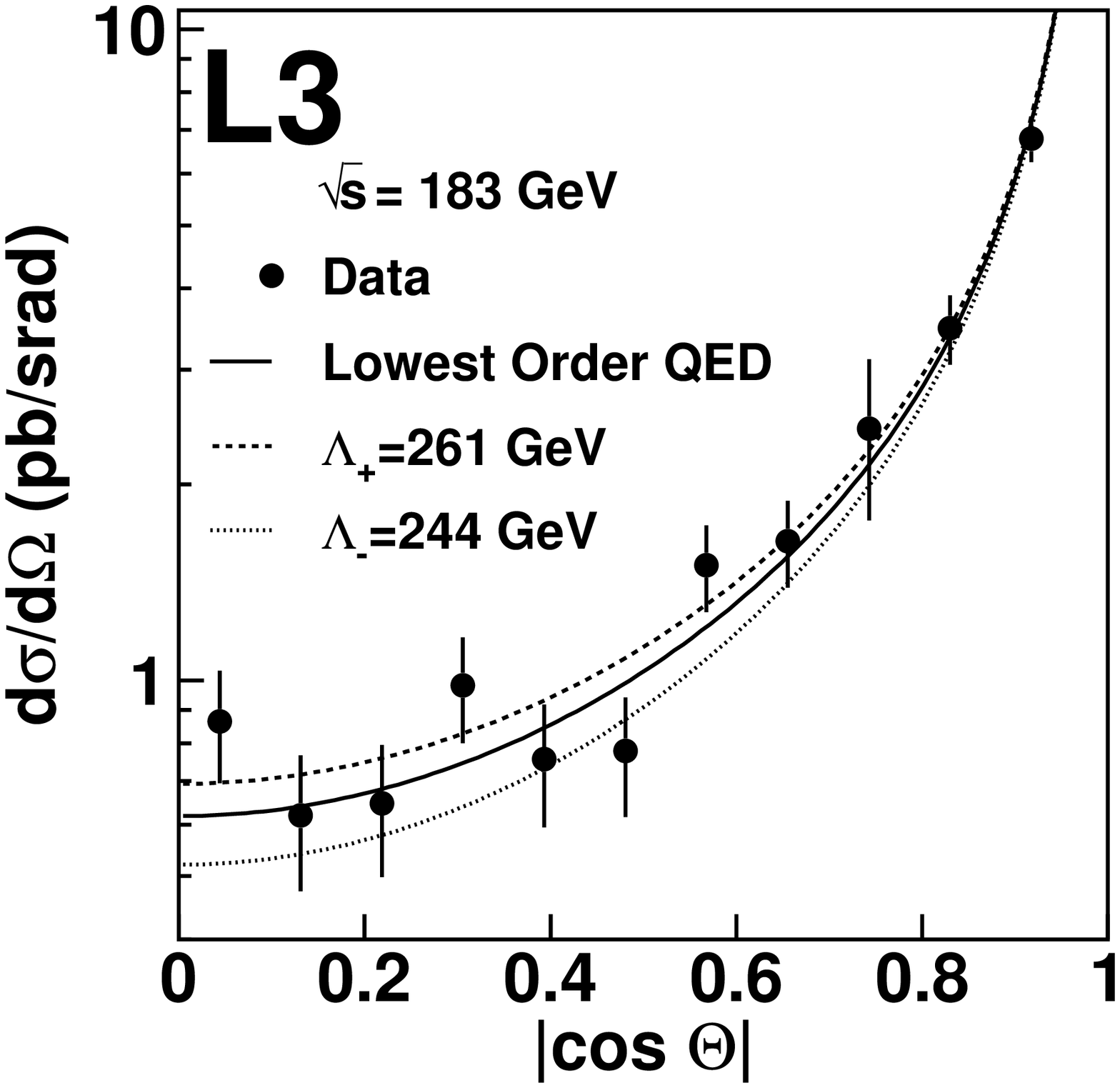}
\end{center}
\begin{center}
\includegraphics[width=10.0truecm]{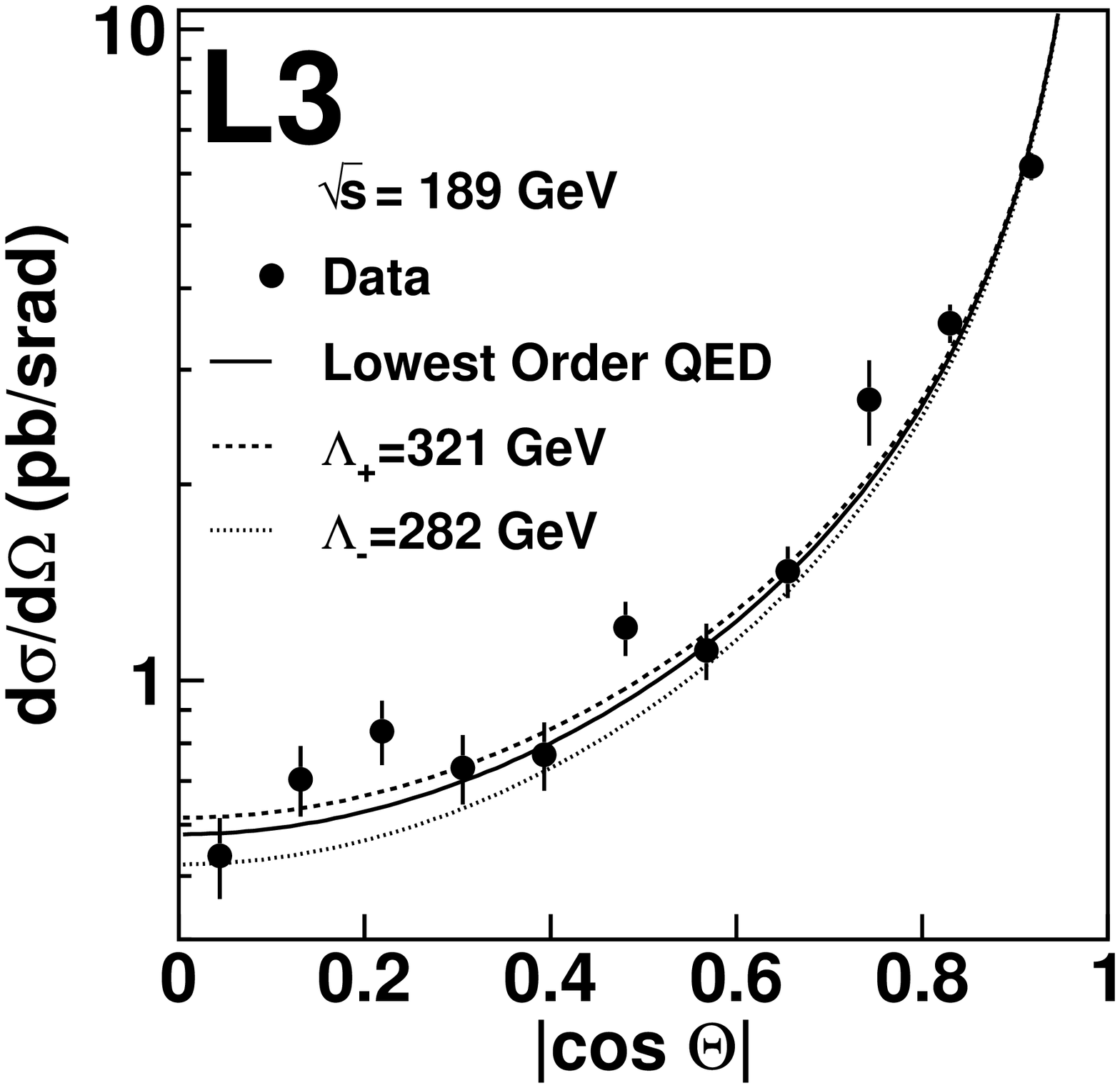}
\end{center}
  
\icaption{\label{fig:costheta}
    Differential cross section as a function of $\cos \theta$ for the process
    $\epem \ra \gamma\gamma(\gamma)$. The points show the measurements
    corrected for efficiency and additional photons. The solid line corresponds to the 
    lowest order QED prediction. The dashed and dotted lines 
    represent the limits obtained for deviations from QED, taking into account 
    all the L3 data at centre-of-mass energies up to that presented 
    in the corresponding plot. }
\end{figure}

The differential cross section as a function of the $\cos \theta$ of the event 
is shown in Figure~\ref{fig:costheta}. The polar angle ${\theta}$ of the 
event is defined as 
${\rm cos}{\theta}= |{\rm sin}(\frac{{\theta}_1-{\theta}_2}{2})/
{\rm sin}(\frac{{\theta}_1+{\theta}_2}{2})|$, where ${\theta}_1$ and ${\theta}_2$ are 
the polar angles of the two most energetic photons in the event. The measured differential
distributions have been corrected for efficiency and higher order 
QED contributions using the Monte Carlo 
simulation. These distributions are then compared 
directly with the lowest order QED predictions. Good agreement between the data 
and the QED prediction is observed.

The observed number of events corresponds to a total cross section in 
the fiducial region $16^{\circ} < \theta < 164^{\circ}$ of:

\vspace*{-0.5truecm}
\begin{eqnarray*}
\sigma_{\gamma\gamma(\gamma)} &=& 12.17 \pm 0.55 \pm 0.14~\mathrm{pb}~~~(\sqrt{s}=183{\rm\GeV{}}) \\
\sigma_{\gamma\gamma(\gamma)} &=& 11.54 \pm 0.30 \pm 0.14~\mathrm{pb}~~~(\sqrt{s}=189{\rm\GeV{}}),
\end{eqnarray*}

\noindent
where the first error is statistical and the second is systematic. The main source of 
systematic error is the uncertainty in the selection efficiency. It has been evaluated
by varying the selection cuts and taking into account the finite Monte Carlo statistics. The
systematic error coming from the uncertainty in the measured luminosity ($\pm 0.2$\%) and 
in the background present in the sample ($< 0.5$\%) are found to be negligible. The 
statistical error dominates in the measurement of the cross section both at $183\GeV$ 
and at $189\GeV$. The QED predicted cross 
sections are 12.65 pb and 11.78 pb~\cite{ggg} respectively, in agreement 
with the measurements. 

\begin{figure}
\begin{center}
\includegraphics[width=18.0truecm]{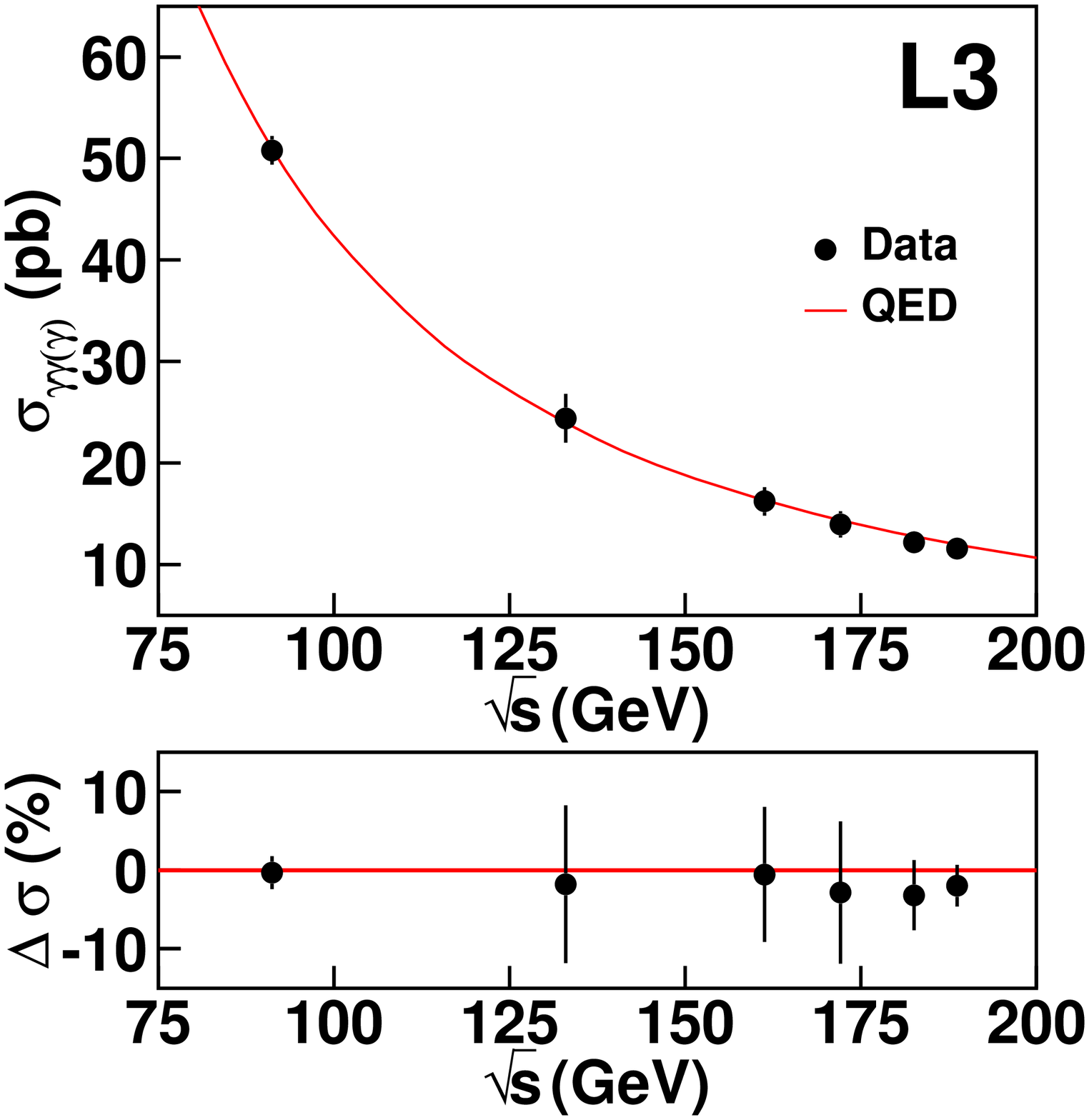}
\vspace{-4.truecm}

\vspace*{-1.0truecm}
\icaption{\label{fig:qedevol}
    Measured cross section as a function of the centre-of-mass energy for
    $\theta$ between $16^{\circ}$ and $164^{\circ}$ compared with the
    QED prediction. The value at $\sqrt{s}=91\GeV$ has been extrapolated
    to this angular range from the one given in {\protect \cite{gg91}}. The bottom
    part of the figure presents the relative deviation of the measurements with respect
    to the QED expectations.}
\end{center}
\end{figure}

These cross sections and previously measured values~\cite{gg91,gg130,gg172} together
with the QED prediction, are presented in Figure~\ref{fig:qedevol} as a function of the
centre-of-mass energy. 

%
%
\section{Limits on deviations from QED}
\indent 

The possible deviations from QED are parametrised by effective Lagrangians, and 
their effect on the observables can be expressed as a multiplicative correction term to 
the QED differential cross section. Depending on the type of Lagrangian, two general 
forms are considered~\cite{lambda}:

\begin{eqnarray}
\frac{d{\sigma}}{d{\Omega}}\,=\,
\left(\frac{d{\sigma}}{d{\Omega}}\right)_{\rm QED}\,
\left(1+\frac{s^2}{\alpha} \frac{1}{{\Lambda}^4} {\rm sin}^2{\theta} \right)
\label{eq:lambda}
\end{eqnarray}

\noindent and

\begin{eqnarray}
\frac{d{\sigma}}{d{\Omega}}\,=\,
\left(\frac{d{\sigma}}{d{\Omega}}\right)_{\rm QED}\,
\left(1+\frac{s^3}{32{\pi}{\alpha}^2} \frac{1}{{\Lambda}^{'6}}
\frac{{\rm sin}^2{\theta}}{1+{\rm cos}^2{\theta}} \right).
\label{eq:lambdaprime}
\end{eqnarray}

\noindent
The correction factors depend on the centre-of-mass energy, the polar angle 
and the scale parameters $\Lambda$, $\Lambda^{'}$ which have dimensions of 
energy. A more standard way of parametrising the deviations from QED is the introduction 
of the cut-off parameters 
${\Lambda}_{\pm}$~\cite{cutoff}. The differential cross section can be obtained from 
equation~(\ref{eq:lambda}) by replacing ${\Lambda}^4$ by ${\pm}(2/{\alpha}){\Lambda}^4_{\pm}$.

Limits on the different scale parameters have already been set in our previous 
publications~\cite{gg130,gg172}. However, since the sensitivity to possible deviations from QED 
increases rapidly with the centre-of-mass energy they are superseded by the present data. In
order to quantify the possible deviations from QED we perform a maximum likelihood fit to 
the differential cross sections at each centre-of-mass energy. The estimated parameters 
combining the present results with those in our previous analyses~\cite{gg130,gg172} are:

\begin{eqnarray}
 \frac{1}{{\Lambda}^4}\, &=& \left(-0.019^{+\, 0.054}
   _{-\, 0.038} \right)\,\times 10^{-11}\,\,\,{\rm\GeV{}}^{-4}   \nonumber \\
 \frac{1}{{{\Lambda}^{'}}^6}\, &=& \left(-0.048^{+\, 0.131}
     _{-\, 0.092} \right)\,\times 10^{-16}\,\,\,{\rm\GeV{}}^{-6} \nonumber
\end{eqnarray}

\noindent
consistent with no deviations from QED. To determine the confidence levels, the probability 
distribution is normalised over the physically allowed range of the parameters. At the 
$95\%$ C.L. the following limits are obtained:

\begin{center}
\begin{tabular}{cc}
 $\Lambda     > 1304\GeV$,   & ${\Lambda}_{+} > 321\GeV$ \\
 $\Lambda^{'} >  703\GeV$,   & ${\Lambda}_{-} > 282\GeV$ \\
\end{tabular}
\end{center}

The effects of $\Lambda_{\pm}$ in the differential cross section can be seen in 
Figure~\ref{fig:costheta}. In this case, the parameters $\Lambda_{\pm}$ have
been fixed to the limits values quoted before.

The existence of excited electrons $({\rm e}^*)$  would also introduce deviations from the QED
predictions in the $\gamma\gamma$($\gamma$) final states. The excited electron, of mass $m_{\rm e^*}$, couples
to ${\rm e}$ and $\gamma$ via two possible interactions. The first is purely 
magnetic~\cite{estr1}, 

\begin{center}
\begin{tabular}{c}
${\cal{L}}\,=\,\frac{\textstyle e}{\textstyle 2{\Lambda}_{\rm e^*}}\,\overline{\Psi}_{\rm e^*} {\sigma}^{\mu\nu}{\Psi}_{\rm e} F_{\mu\nu}\,+\, h.c.$  \\
\end{tabular}
\end{center}

\noindent
and the second is a chiral-magnetic one~\cite{estr2}:

\begin{center}
\begin{tabular}{c}
${\cal{L}}\,=\,\frac{\textstyle e}{\textstyle 2{\Lambda}_{\rm e^*}}\,\overline{\Psi}_{\rm e^*} {\sigma}^{\mu\nu} (1{\pm}{\gamma}^5){\Psi}_{\rm e} F_{\mu\nu}\,+\, h.c.$
\end{tabular}
\end{center}

\noindent
In both cases we fit the excited electron mass fixing the interaction 
scale ${\Lambda}_{\rm e^*}$ to $m_{\rm e^*}$, obtaining

\begin{center}
\begin{tabular}{lc}
Purely Magnetic: &  $\frac{\textstyle 1}{\textstyle m_{\rm e^*}^4} = \left(-0.052^{+\, 0.143}_{-\, 0.104}\right)\,\times 10^{-9}\GeV^{-4}$  \\ 
Chiral-Magnetic: &  $\frac{\textstyle 1}{\textstyle m_{\rm e^*}^4} \, = \left(-0.135^{+\, 0.383}_{-\, 0.352} \right)\,\times 10^{-9}\GeV^{-4}$   
\end{tabular}
\end{center}

\noindent
From them we derive the $95\%$ C.L. lower limits of:

\begin{center}
\begin{tabular}{lc}
Purely Magnetic: &  $m_{\rm e^*} > 283\GeV$ \\
Chiral-Magnetic: &  $m_{\rm e^*} > 213\GeV$
\end{tabular}
\end{center}

%
%
\section{Acknowledgements}
\indent

We wish to express our gratitude to the CERN accelerator divisions
for the excellent performance of the LEP machine. We acknowledge
the effort of the engineers and technicians who have participated
in the construction and maintenance of the experiment.

%
%

%

\typeout{   }     
\typeout{Using author list for paper 195 -?}
\typeout{$Modified: Tue Nov 23 09:52:18 1999 by clare $}
\typeout{!!!!  This should only be used with document option a4p!!!!}
\typeout{   }
%
%
%
%
%
%

\newcount\tutecount  \tutecount=0
\def\tutenum#1{\global\advance\tutecount by 1 \xdef#1{\the\tutecount}}
\def\tute#1{$^{#1}$}
\tutenum\aachen            
\tutenum\nikhef            
\tutenum\mich              
\tutenum\lapp              
\tutenum\basel             
\tutenum\lsu               
\tutenum\beijing           
\tutenum\berlin            
\tutenum\bologna           
\tutenum\tata              
\tutenum\ne                
\tutenum\bucharest         
\tutenum\budapest          
\tutenum\mit               
\tutenum\debrecen          
\tutenum\florence          
\tutenum\cern              
\tutenum\wl                
\tutenum\geneva            
\tutenum\hefei             
\tutenum\seft              
\tutenum\lausanne          
\tutenum\lecce             
\tutenum\lyon              
\tutenum\madrid            
\tutenum\milan             
\tutenum\moscow            
\tutenum\naples            
\tutenum\cyprus            
\tutenum\nymegen           
\tutenum\caltech           
\tutenum\perugia           
\tutenum\cmu               
\tutenum\prince            
\tutenum\rome              
\tutenum\peters            
\tutenum\salerno           
\tutenum\ucsd              
\tutenum\santiago          
\tutenum\sofia             
\tutenum\korea             
\tutenum\alabama           
\tutenum\utrecht           
\tutenum\purdue            
\tutenum\psinst            
\tutenum\zeuthen           
\tutenum\eth               
\tutenum\hamburg           
\tutenum\taiwan            
\tutenum\tsinghua          
{
\parskip=0pt
\noindent
{\bf The L3 Collaboration:}
\ifx\selectfont\undefined
 \baselineskip=10.8pt
 \baselineskip\baselinestretch\baselineskip
 \normalbaselineskip\baselineskip
 \ixpt
\else
 \fontsize{9}{10.8pt}\selectfont
\fi
\medskip
\tolerance=10000
\hbadness=5000
\raggedright
\hsize=162truemm\hoffset=0mm
\def\r{\rlap,}
\noindent

M.Acciarri\r\tute\milan\
P.Achard\r\tute\geneva\ 
O.Adriani\r\tute{\florence}\ 
M.Aguilar-Benitez\r\tute\madrid\ 
J.Alcaraz\r\tute\madrid\ 
G.Alemanni\r\tute\lausanne\
J.Allaby\r\tute\cern\
A.Aloisio\r\tute\naples\ 
M.G.Alviggi\r\tute\naples\
G.Ambrosi\r\tute\geneva\
% ...
\newpage
\begin{list}{A}{\itemsep=0pt plus 0pt minus 0pt\parsep=0pt plus 0pt minus 0pt
                \topsep=0pt plus 0pt minus 0pt}
\item[\aachen]
 I. Physikalisches Institut, RWTH, D-52056 Aachen, FRG$^{\S}$\\
 III. Physikalisches Institut, RWTH, D-52056 Aachen, FRG$^{\S}$
\item[\nikhef] National Institute for High Energy Physics, NIKHEF, 
     and University of Amsterdam, NL-1009 DB Amsterdam, The Netherlands
\item[\mich] University of Michigan, Ann Arbor, MI 48109, USA
\item[\lapp] Laboratoire d'Annecy-le-Vieux de Physique des Particules, 
     LAPP,IN2P3-CNRS, BP 110, F-74941 Annecy-le-Vieux CEDEX, France
\item[\basel] Institute of Physics, University of Basel, CH-4056 Basel,
     Switzerland
\item[\lsu] Louisiana State University, Baton Rouge, LA 70803, USA
\item[\beijing] Institute of High Energy Physics, IHEP, 
  100039 Beijing, China$^{\triangle}$ 
\item[\berlin] Humboldt University, D-10099 Berlin, FRG$^{\S}$
\item[\bologna] University of Bologna and INFN-Sezione di Bologna, 
     I-40126 Bologna, Italy
\item[\tata] Tata Institute of Fundamental Research, Bombay 400 005, India
\item[\ne] Northeastern University, Boston, MA 02115, USA
\item[\bucharest] Institute of Atomic Physics and University of Bucharest,
     R-76900 Bucharest, Romania
\item[\budapest] Central Research Institute for Physics of the 
     Hungarian Academy of Sciences, H-1525 Budapest 114, Hungary$^{\ddag}$
\item[\mit] Massachusetts Institute of Technology, Cambridge, MA 02139, USA
\item[\debrecen] KLTE-ATOMKI, H-4010 Debrecen, Hungary$^\P$
\item[\florence] INFN Sezione di Firenze and University of Florence, 
     I-50125 Florence, Italy
\item[\cern] European Laboratory for Particle Physics, CERN, 
     CH-1211 Geneva 23, Switzerland
\item[\wl] World Laboratory, FBLJA  Project, CH-1211 Geneva 23, Switzerland
\item[\geneva] University of Geneva, CH-1211 Geneva 4, Switzerland
\item[\hefei] Chinese University of Science and Technology, USTC,
      Hefei, Anhui 230 029, China$^{\triangle}$
\item[\seft] SEFT, Research Institute for High Energy Physics, P.O. Box 9,
      SF-00014 Helsinki, Finland
\item[\lausanne] University of Lausanne, CH-1015 Lausanne, Switzerland
\item[\lecce] INFN-Sezione di Lecce and Universit\'a Degli Studi di Lecce,
     I-73100 Lecce, Italy
\item[\lyon] Institut de Physique Nucl\'eaire de Lyon, 
     IN2P3-CNRS,Universit\'e Claude Bernard, 
     F-69622 Villeurbanne, France
\item[\madrid] Centro de Investigaciones Energ{\'e}ticas, 
     Medioambientales y Tecnolog{\'\i}cas, CIEMAT, E-28040 Madrid,
     Spain${\flat}$ 
\item[\milan] INFN-Sezione di Milano, I-20133 Milan, Italy
\item[\moscow] Institute of Theoretical and Experimental Physics, ITEP, 
     Moscow, Russia
\item[\naples] INFN-Sezione di Napoli and University of Naples, 
     I-80125 Naples, Italy
\item[\cyprus] Department of Natural Sciences, University of Cyprus,
     Nicosia, Cyprus
\item[\nymegen] University of Nijmegen and NIKHEF, 
     NL-6525 ED Nijmegen, The Netherlands
\item[\caltech] California Institute of Technology, Pasadena, CA 91125, USA
\item[\perugia] INFN-Sezione di Perugia and Universit\'a Degli 
     Studi di Perugia, I-06100 Perugia, Italy   
\item[\cmu] Carnegie Mellon University, Pittsburgh, PA 15213, USA
\item[\prince] Princeton University, Princeton, NJ 08544, USA
\item[\rome] INFN-Sezione di Roma and University of Rome, ``La Sapienza",
     I-00185 Rome, Italy
\item[\peters] Nuclear Physics Institute, St. Petersburg, Russia
\item[\salerno] University and INFN, Salerno, I-84100 Salerno, Italy
\item[\ucsd] University of California, San Diego, CA 92093, USA
\item[\santiago] Dept. de Fisica de Particulas Elementales, Univ. de Santiago,
     E-15706 Santiago de Compostela, Spain
\item[\sofia] Bulgarian Academy of Sciences, Central Lab.~of 
     Mechatronics and Instrumentation, BU-1113 Sofia, Bulgaria
\item[\korea] Center for High Energy Physics, Adv.~Inst.~of Sciences
     and Technology, 305-701 Taejon,~Republic~of~{Korea}
\item[\alabama] University of Alabama, Tuscaloosa, AL 35486, USA
\item[\utrecht] Utrecht University and NIKHEF, NL-3584 CB Utrecht, 
     The Netherlands
\item[\purdue] Purdue University, West Lafayette, IN 47907, USA
\item[\psinst] Paul Scherrer Institut, PSI, CH-5232 Villigen, Switzerland
\item[\zeuthen] DESY, D-15738 Zeuthen, 
     FRG
\item[\eth] Eidgen\"ossische Technische Hochschule, ETH Z\"urich,
     CH-8093 Z\"urich, Switzerland
\item[\hamburg] University of Hamburg, D-22761 Hamburg, FRG
\item[\taiwan] National Central University, Chung-Li, Taiwan, China
\item[\tsinghua] Department of Physics, National Tsing Hua University,
      Taiwan, China
\item[\S]  Supported by the German Bundesministerium 
        f\"ur Bildung, Wissenschaft, Forschung und Technologie
\item[\ddag] Supported by the Hungarian OTKA fund under contract
numbers T019181, F023259 and T024011.
\item[\P] Also supported by the Hungarian OTKA fund under contract
  numbers T22238 and T026178.
\item[$\flat$] Supported also by the Comisi\'on Interministerial de Ciencia y 
        Tecnolog{\'\i}a.
\item[$\sharp$] Also supported by CONICET and Universidad Nacional de La Plata,
        CC 67, 1900 La Plata, Argentina.
\item[$\diamondsuit$] Also supported by Panjab University, Chandigarh-160014, 
        India.
\item[$\triangle$] Supported by the National Natural Science
  Foundation of China.
\item[\dag] Deceased.
\end{list}
}
\vfill







\newpage


\newpage


\newpage


\newpage


\newpage

%
\end{document}